\documentclass[conference,a4paper]{IEEEtran}
\usepackage[pdftex]{graphicx}
\usepackage{amsmath,amssymb}

\def\(({\left(}
\def\)){\right)}
\def\[[{\left[}
\def\]]{\right]}

\newcommand{\be}{\begin{equation}}
\newcommand{\ee}{\end{equation}}
\newcommand{\bea}{\begin{eqnarray}}
\newcommand{\eea}{\end{eqnarray}}

\newcommand{\calF}{\mathcal F}

\begin{document}

\sloppy

\title{Phase Diagram and Approximate Message Passing for Blind
  Calibration and Dictionary Learning}

\author{\IEEEauthorblockN{Florent Krzakala}
\IEEEauthorblockA{
ESPCI and CNRS UMR 7083 \\
10 rue Vauquelin,\\
Paris 75005  France\\
fk@espci.fr}
\and
\IEEEauthorblockN{Marc M\'ezard}
\IEEEauthorblockA{Ecole Normale Sup\'erieure\\ 45 rue d'Ulm, Paris
  France\\ 
and LPTMS-CNRS Univ. Paris Sud \\
Orsay, France}
\and
\IEEEauthorblockN{Lenka   Zdeborov\'a}
\IEEEauthorblockA{Institut de Physique Th\'eorique\\ IPhT, CEA Saclay\\ and URA 2306,
CNRS\\ 91191 Gif-sur-Yvette, France.}}


\maketitle

\begin{abstract}
  We consider dictionary learning and blind calibration for signals
  and matrices created from a random ensemble. We study the
  mean-squared error in the limit of large signal dimension using the replica method and
  unveil the appearance of phase transitions delimiting impossible,
  possible-but-hard and possible inference regions. We also introduce
  an approximate message passing algorithm that asymptotically matches
  the theoretical performance, and show through numerical tests that
  it performs very well, for the calibration problem, for tractable
  system sizes.
\end{abstract}

\section{Introduction}
Matrix decomposition $Y=FX$, where $Y$ is known and one seeks $F$ and $X$, with requirements (e.g. sparsity, probability distribution of
elements etc.) on the properties of $X$ and $F$, is a generic problem
that appears in many applications \cite{IgorBlog}. Theoretical limits
on when matrix decompositions is possible and tractable are still very
poorly understood. In this work we make a step towards
this understanding by determining the limits of matrix decomposition
when $Y$ is created using randomly generated matrices $F$ and $X$.

Consider a set of $P$ $K$-sparse $N$-dimensional vectors (``signals''),
with iid components created from the distribution (denoting $\rho=K/N$) \be P(x^0_{il})\!
= \!(1-\rho) \delta(x^0_{il}) + \rho {\phi} (x^0_{il}), \label{Px} \ee
where $i\!=\!1,\ldots,N$, $l\!=\!1,\ldots,P$. For each of the vectors
we perform $M$ linear measurements, summarized by a $M\!  \times\!  N$
measurement matrix with iid elements $\calF^0_{\mu i}=F^0_{\mu
  i}/\sqrt{N}$ (this rescaling ensures that the measurements are of $O(1)$) where the $F^0_{\mu i}$ are generated from a
probability distribution $\phi_F(F^0_{\mu i})$ (in the numerical
examples we will always consider a Gaussian distribution of zero mean
and unit variance). We have only access to the (noisy) results of
these measures, that is, to the $P$ vectors ${y}_l$ such
that
 \be y_{\mu l\!}=\!\sum_{i} \calF_{\mu i}^0 x_{il}^0+ \xi_{\mu l} \, , \ee 
where $\xi_{\mu l}$ is a Gaussian additive
noise with variance $\Delta$.

Is it possible to find both the vectors $x^0$ and the matrix
(dictionary) $F^0$ (up to a permutation of $N\!$ elements and their
signs)? This is the {\it dictionary learning} problem. A related
situation is when one knows at least a noisy version of the matrix
$F^0$, defined by $F'=(F^0 + \sqrt{\eta} W)/\sqrt{1+\eta}$ where $W$ is a random
matrix with the same statistics as $F^0$.  $P(F^0|F')$ then reads, for each matrix element 
\be
P(F^0_{\mu i}| F'_{\mu i} ) = {\cal N}(\frac{F'_{\mu
    i}}{\sqrt{1+\eta}}, \frac{\eta}{1+\eta}) \, .  \label{PF}\ee 
Recovering $F^0$ and $x^0$, knowing this time $F'$ and the $P$
vectors $y_{l}$ is a problem that we shall refer to as {\it blind
  calibration}. It becomes equivalent to dictionary learning when
$\eta \to \infty$. 

Our goal here is to analyse optimal Bayes inference (that provides the MMSE
(minimal MSE)) where the signal $x_{il}^0$ and the dictionary
$F^0_{\mu i}$ are estimated from the marginals of the posterior
probability
\bea 
&&P(x_{il},F_{\mu i} | y_{\mu l}, F'_{\mu i}) = 
\frac{1}{Z} \prod_{\mu i} P(F_{\mu i}|F'_{\mu
  i})  \nonumber \\ &&   \prod_{il} \!P(x_{il}) \prod_{\mu l} \!\! \left[ \frac{1}{\sqrt{2\pi
      \Delta}} e^{-\frac{(y_{\mu l} - \sum_i F_{\mu
      i}x_{il}/\sqrt N )^2}{2\Delta}} \! \right]
\label{Bayes} \, .\eea

\subsection{Related works} 
There are several algorithm suggested and tested for dictionary
learning, see e.g.
\cite{OlshausenField97,EnganAase99,AharonElad06b,MairalBach09}. The algorithm we derive in this paper is 
closely related but different to the bilinear AMP proposed by \cite{SCHNITER-BIG} as
explained in Sec.~\ref{secBP}. 

The question of how many samples $P$ are necessary
for the dictionary to be identifiable has a straightforward lower
bound $MP > N(M + P \rho)$, otherwise there is more unknown variables
than measurements and hence exact recovery is clearly
impossible. Several works analyzed what is a sufficient number of
samples for exact recovery. While early rigorous results were able to
show learnability from only exponential many samples
\cite{AharonElad06}, more recent analysis of convex relaxation based
approaches shows that $O(N\log{N})$ samples are needed for $\alpha=1$
\cite{GribonvalSchnass09} and polynomially many for $\alpha<1$
\cite{GengWangy11}. Another study of sample complexity for dictionary
learning for $\alpha<1$ establishes a $O(N\log{N})$ bound for the
number of samples \cite{Vainsencher11}. A very recent non-rigorous
work suggested that $P\!=\!O(N)$ samples should be sufficient to
identify the dictionary \cite{SakataKabashima12}. That work was based
on the replica analysis of the problem, but did not analyze the
Bayes-optimal approach. 

Several works also considered blind calibration, where only a
uncertain version of the matrix $F$ is known, on the other hand one
has the access to many signals and their measurements such that
calibration of the matrix $F$ is possible, see
e.g. \cite{GribonvalChardon11} and reference therein. Cases when both
the signal and the dictionary are sparse are also considered in the
literature, e.g. \cite{Rubinstein10}, and our theory can be applied to
these as well.

\subsection{Main results} 
The present paper has three main results. First, using the replica
method \cite{MezardParisi87b} we estimate the Bayes
optimal MMSE in the limit of large signals $N \to \infty$. In
particular, we define $\alpha\!=\!M/N$, $\pi\!=\!P/N$ and show that
for the noiseless case, $\Delta=0$, exact reconstruction is possible
if $\pi\!>\!\pi^*\!=\!{\alpha}/{(\alpha-\rho)}$ and $\alpha>\rho$.  In
this regime, it is thus possible to recover the matrix and the signal
exactly if one can compute the marginals of the posterior probability
distribution (\ref{Bayes}). This result is striking, all the more
because it is independent of $\eta$.

Computing the marginals of the posterior probability distribution is
an extremely hard problem, all the more when there is a phase
transition in the problem. We determine the value $\pi^s(\eta)$ (the
spinodal transition) below which iterative sampling (using for
instance Monte Carlo Markov chains or message passing) is believed to
be intractable in polynomial time.

Finally, we introduce an AMP-like message passing algorithm designed to
perform such sampling, and show that it performs very well for the
calibration problem. However, at the moderate size we are able to
study numerically, finite-size deviation from the asymptotic behavior
become large as $\eta$ grows and prevents our algorithm to function
nicely in the dictionary learning limit. However, we believe that this
still sets a very promising stage for new algorithmic development for
dictionary learning.

\section{Asymptotic analysis with the replica method}
\label{sec:replica}

\subsection{Replica analysis for matrix decomposition} 
\begin{figure}[t]
\centering
\includegraphics[width=3.2in]{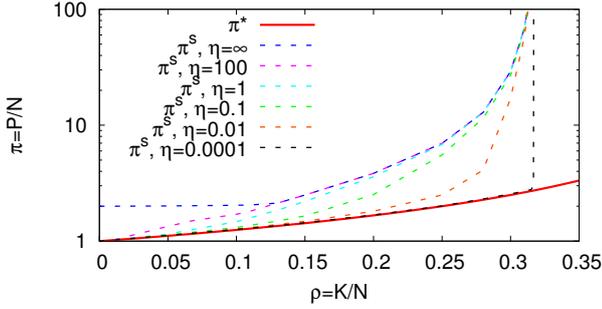}
\caption{Phase diagram for $\alpha\!=\!0.5$ and $\Delta\!=\!0$. For
  both blind calibration and dictionary learning, exact learning is
  possible by the Bayes optimal approach above the full red line
  $\pi^*=\frac{\alpha}{\alpha-\rho}$. However, such a sampling
  procedures will not be tractable below the spinodal transition
  $\pi^s(\eta)$ shown here in dotted line for dictionary learning
  ($\eta\!=\!\infty$) and blind calibration.}
\label{fig1}
\end{figure}
\begin{figure}[!ht]
\centering
\includegraphics[width=3.2in]{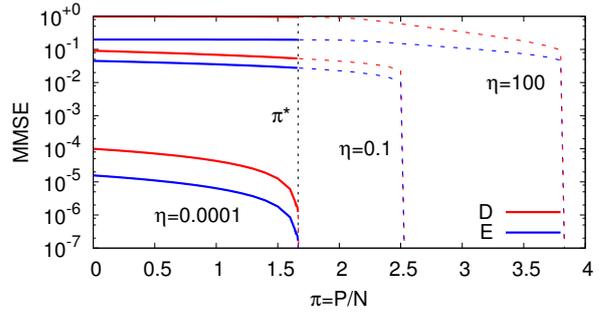}
\caption{MMSE $D$ (for the matrix $F$) and $E$ (for the signal
  ${x}$) corresponding to $\rho=0.2$, $\alpha=0.5$, $\Delta=0$, for
  three values of $\eta$. The MMSE jumps
  abruptly from a finite value to zero at $\pi^*$.  However, sampling
  should remain intractable until the spinodal transition arises at a
  larger value $\pi^s(\eta)$ and we denote the corresponding MSE in
  dotted line. The figure remains qualitatively the same for small
  value of the additive noise $\Delta>0$, where the MMSE at large
  $\pi$ is not zero but rather $O(\Delta)$. If $\Delta$ is large
  enough, however, the sharp transition disappears and the MMSE is
  continuous (see e.g. Fig.~\ref{bpplot}).}
\label{fig_MMSE}
\end{figure}

The MMSE obtained from the Bayes-optimal approach can be computed exactly in
the limit of large $N$ via the replica method. Although this method is
in general non-rigorous, it is sometimes possibles to prove that the
results derived from it are exact.,
We shall leave out details of the derivation and refer instead
to~\cite{KrzakalaPRX2012,KrzakalaMezard12} for a very similar
computation in the case of compressed sensing.  We estimate the Bayes
optimal MMSE by computing the marginals of the matrix and signals elements. Our
  computation is also very similar to the one of
  \cite{SakataKabashima12}, who however did not analyze the Bayes
  optimal MMSE.

  We now compute $ \Phi= {\mathbb
    E}_{F^0,F',x^0}(\log{Z(F^0,F',x^0)})/NP$, where $Z$, the so-called
  partition sum (or evidence), is the normalization in
  eq.~(\ref{Bayes}) for a given instance of the problem. In order to
  do so we compute ${\mathbb E}_{F^0,F',x^0}(Z^n)$ and then use the
  replica trick $\log{Z}=\lim_{n\to 0} (Z^n-1)/n$. We use the replica
  symmetric ansatz which is correct for inference problems with prior
  distributions corresponding to the generative process.  The final
  result is that in the large signal limit, $N\to \infty$, the MMSE
  $D$ (on the matrix) and $E$ (on the signals) are  given by
\be
{\rm MMSE}(\alpha,\pi,\rho,\Delta,\eta)=\arg\!\max_{E,D} \Phi(E,D) \, ,
\ee
where the so-called ``potential'' is given by
 \bea
&&\!\!\!\!\!\!\!\!\! \!\!\!\! \Phi(\!E\!,\!D\!)\!\!= \!\! - \frac{\alpha}{2}
\log{(\Delta\!+\!E\!+\!D(\rho-E))} \!\! - \!\!
\frac{\alpha(\Delta+\rho)}{\Delta+E+ D(\rho-E)} \nonumber\\
&+&\!\!\!\! \frac{\alpha}{2}+\!\!  \[[\int {\cal D}z \log{ \left\{ \[[
    e^{-\frac{\hat m_x}{2} x^2 +\hat m_x x x^0 + z \sqrt{\hat m_x}x
    } \]]_{P(x)} \right\} }\]]_{P(x^0)} \nonumber\\ &+& \!\!\!\!\!\!
\frac{\alpha}{\pi} \!\!
\[[  \int    {\cal      D}z \! \log{ 
  \[[ e^{-\frac{\hat m_F F^2}{2}+\hat m_F F F^0 + z \sqrt{\hat m_F}F
  } \]]_{P(F|F')} }\]]_{\!P(F^0\!,F')} \, .\nonumber \eea 
Here $[f(u)]_{Q(u)}$ denotes an average of a function $f$ of the
random variable $u$ with distribution $Q(u)$, $ {\cal D}z$ a Gaussian
measure with zero mean and unit variance. The probability
distributions $P(x)$, and $P(F|F')$ are the single-element
distributions introduced in eqs.~(\ref{Px}) and (\ref{PF}). 
Finally we denoted 
\be
\hat m_x \!=\! \frac{\alpha (1-D)}{ \Delta + E + \rho D - E D },  ~~~
\hat m_F \!=\! \frac{\pi (\rho-E )}{ \Delta + E + \rho D - E D } \, . \label{hat_eq}
\ee
Note that the present expression for the potential is very general and
can be used to study many similar problems, such as matrix completion,
or sparse matrix decomposition, by changing the distribution of the
matrix and of the signal.

\subsection{Gaussian matrix and Gauss-Bernoulli signal}
When the matrix elements $F^0_{\mu i}$ are generated from a
Gaussian with zero mean and unit variance, and $x^0_{il}$ from
Gauss-Bernoulli distribution, the potential simplifies to
\bea && 
\!\!\!\!\!\!\!\!\! \!\!\!\! \Phi(E,D) \!\!=\!\!-\frac{\alpha}{2} \!
\log{(\Delta\!+\!E\!+\! D(\rho-E))} \!\! - \!\!
\frac{\alpha(\Delta+\rho)}{\Delta\!+\!E\!+\!D(\rho-E)} \nonumber\\
&+& \!\! \frac{\alpha}2 + (1-\rho) \int
{\cal D}z \, \log{\left( 1 -\rho + \frac{\rho}{\sqrt{\hat m_x +1}}
    e^{\frac{z^2\hat m_x}{2(\hat m_x +1)}} \right)}  \nonumber \\ 
&+& \!\!\!\rho \! \int {\!\! \cal
  D}z \log{\!\left(\!1\! -\!\rho \!+\! \frac{\rho}{\sqrt{\hat m_x\!+\!1}}
    e^{\frac{z^2\hat m_x}{2}}\! \right)} + \frac{\alpha}{2\pi}\! \hat m_F \nonumber \\
&-& \frac{\alpha}{2\pi} \log{\left(1+\frac{\eta \hat m_F
    }{1+\eta}\right)} \, . \label{replica}
\eea
The Bayes-optimal MMSE is obtained by maximizing $\Phi(E,D)$.  
Analyzing the above expression in the zero-noise limit ($\Delta = 0$)
allows to demonstrate our first main result: in both the blind
calibration and dictionary learning problems, the global maximum is
given by $D\! =\!E\!=\!0$, with $\Phi \to \infty$, as long as
$\alpha\!>\!\rho$ and $\pi\!>\!\pi^*\!=\!{\alpha}/{(\alpha-\rho)}$.
Hence for $\pi\!>\!\pi^*$ it is possible to learn the matrix
and the signal exactly from the measurements. 
This result is striking since it is independent of $\eta$ as long as $\eta>0$; and
coincides with the simple counting lower bound. 
When $\eta \to 0$, more precisely when $\eta \ll \Delta$, then the
compressed sensing phase transition ---that goes to $\alpha=\rho$  when
$\Delta \to 0$--- is recovered independently of $\pi$.

Bayes-optimal learning, however, requires exact sampling from the
measure (\ref{Bayes}), and this remains an extremely hard
computational problem. In this regard, another important transition
(the ``spinodal''), that can be studied from the form of the potential
function $\Phi(E,D)$, is the appearance of a high-MSE local
maxima. This phenomenon marks the downfall of many sampling strategy,
such as Gibbs sampling or message-passing strategy (that 
performs a steepest ascent in the $\Phi(E,D)$ function). We determine
the value $\pi^s(\eta)$ (the so-called ``spinodal'' transition) above
which $\Phi(E,D)$ does not have the spurious secondary maxima. As shown in
Fig.~\ref{fig1} and Fig.~\ref{fig_MMSE}, it depends strongly on the
value $\eta$. In fact, when $\eta \to 0$, we recover again the
compressed sensing limit we have obtained in~\cite{KrzakalaMezard12}.
Figs.~\ref{fig_MMSE} and \ref{bpplot} depict the values of MMSE for the signal $E$ and
the matrix $D$ reached for large systems $N\to \infty$ by the Bayes
optimal sampling and by local sampling strategies 
that reach $D=E=0$ discontinuously at $\pi^s$.

The behavior with finite noise $\Delta<\eta$ is also interesting and
we observe a phenomenology similar to that described
in~\cite{KrzakalaMezard12} in the case of compressed sensing. Because
of a two-maxima shape of the function $\Phi(E,D)$ for moderate
$\Delta$, the MMSE displays a sharp transition separating a region of
parameters with a small MMSE, comparable to $\Delta$, from a region
with a larger $O(\eta)$ MMSE. For larger value of $\Delta$ we do not
see any abrupt transition, and the MMSE continuously decays with
$\pi$ (see e.g. Fig.~\ref{bpplot}).

To conclude, there exist three regions in the phase diagram,
corresponding to impossible (below $\pi^*$), intractable (below
$\pi^s$), and perhaps-tractable learning. We will now study
algorithmically the later one.

\section{Message-passing algorithm}
\label{secBP}
To make the Bayes optimal sampling tractable we have to resort to
approximations. In compressed sensing, the Bayesian approach combined
with a belief propagation (BP)
reconstruction algorithm leads to the so-called approximate message
passing (AMP) algorithm. It was first derived in \cite{DonohoMaleki09}
for the minimization of $\ell_1$, and subsequently generalized in
\cite{DonohoMaleki10,Rangan10b}. We shall now adapt
this strategy to the present case. 

\subsection{From BP to AMP} 

\begin{figure}[!t]
\centering
\includegraphics[width=3.2in]{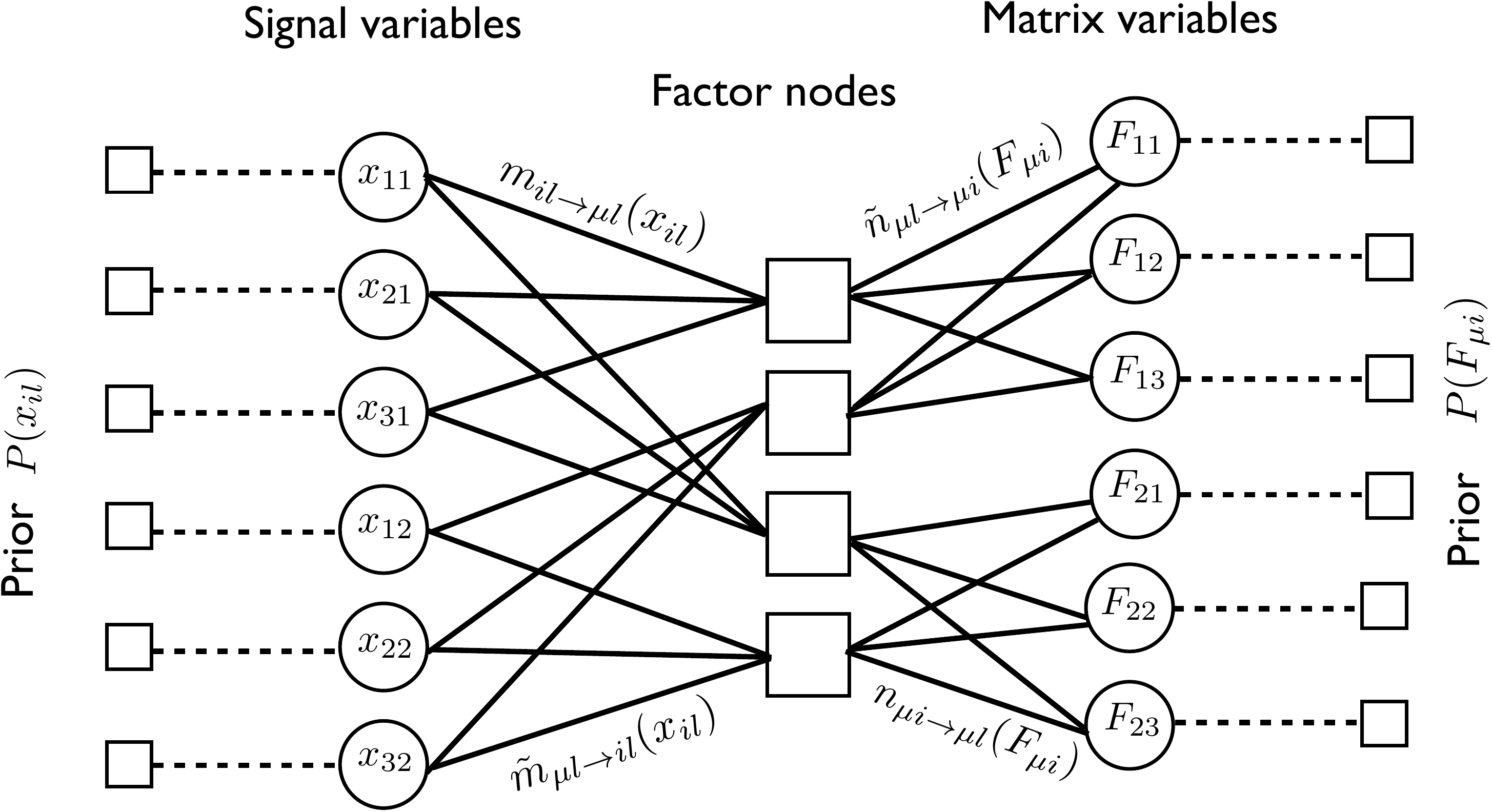}
\caption{Factor graph used for the belief propagation inference, here
  drawn using $N=3$, $P=2$ and $M=2$. The factor nodes ensure (in probability) the  condition $y_{\mu l\!}=\!\sum_{i} \calF_{\mu i} x_{il} +
\xi_{\mu l}$. }
\label{FactorGraph}
\end{figure}

The factor graph corresponding to the posterior
probability (\ref{Bayes}) is depicted in Fig.~\ref{FactorGraph}. The
canonical BP iterative equations are written for
messages $m,n,\hat m,\hat n$ and read  
\bea m_{il\to\mu l} (x_{il})
&\propto& P(x_{il}) \prod_{\nu \neq \mu}^M
\hat{m}_{\nu l \to  i l} (x_{il}) \, ,\\
n_{\mu i \to\mu l} (F_{\mu i}) &\propto&P(F_{\mu i}|F'_{\mu i}) \prod_{n \neq
  l}^P
\hat{n}_{\mu n  \to  \mu i} (F_{\mu i }) \, ,\\
\hat{m}_{\mu l \to i l} (x_{il}) &\propto& \int \prod_{j\neq i}
dx_{jl} dF_{\mu k} e^{-\frac{(y_{\mu l} - \sum_i F_{\mu i}x_{il}/\sqrt
    N )^2}{2\Delta}}
\nonumber \\
&&\prod_k n_{\mu k \to \mu l} (F_{\mu k}) \prod_{j \neq i} m_{jl \to
  \mu l} (x_{jl}) \, ,\\
\hat{n}_{\mu l \to \mu i} (F_{\mu i}) &\propto& \int dx_{jl} \prod_{k
  \neq i} dF_{\mu k}
e^{-\frac{(y_{\mu l} - \sum_i F_{\mu i}x_{il}/\sqrt N )^2}{2\Delta}} \nonumber\\
&&\prod_{k\neq i} n_{\mu k \to \mu l} (F_{\mu k}) \prod_{j} m_{jl \to
  \mu l} (x_{jl}) \, .\eea

A major simplification of these iterative equations arises when one
uses the central limit theorem and realizes that only the two first
moments of the above distributions are important for the leading
contribution when $N\to \infty$. This "Gaussian" approximation is at
the basis of approximate message passing as used in compressed
sensing. The next step of the derivation again neglects $O(1/N)$ terms
and allows to reduce the number of messages to be iterated from
$O(N^4)$ to $O(N^2)$. This leads to a TAP-like set of equations
\cite{ThoulessAnderson77}. Finally if the matrix $F^0$ and signal
$x^0$ have known distribution of elements this further simplifies the
algorithm. A full derivation will be given elsewhere and we instead
refer the reader to the re-derivation of AMP in
\cite{KrzakalaMezard12} where we followed essentially the very same
steps.

The final form of our algorithm for blind calibration and dictionary
learning follows. We denote the mean and variance of the BP estimates
of marginals over $x_{il}$ as $a_{il}$ and $c_{il}$, and those over
$\calF_{\mu i}$ as $r_{\mu i}$ and $s_{\mu l}$. We define several
auxiliary values (all of them are of order $O(1)$): \bea
\overline{a^2} &=& \frac{1}{NP} \sum_{jl} a^2_{jl}, ~~ \overline{c} =
\frac{1}{NP} \sum_{jl} c_{jl} \, ,\nonumber\\
\overline{r^2} &=& \frac{1}{M} \sum_{\nu j} r^2_{\nu j},
~~\overline{s} = \frac{1}{M} \sum_{\nu j} c_{\nu j} \, ,\nonumber \\
\overline{(y-\omega)^2} &=& \frac{1}{MP} \sum_{\nu l} (y_{\nu l} -
\omega_{\nu l})^2\, . \nonumber \eea
Then our AMP algorithm reads
\bea
 \omega_{\nu l }^{t+1} &=& \sum_j r_{\nu j} a_{j l}^t - \frac{y_{\nu
       l}-\omega^t_{\nu l}}{\overline{(y-\omega)^2}^{t}} (  \overline{c}^t
   \overline{r^2}^t+  \overline{a^2}^t \overline{s}^t ) \,\!\! ,\\
   (\Sigma_R^{t+1})^2 &=& \left[\alpha
     \frac{\overline{r^2}^t}{\overline{(y-\omega)^2}^{t+1}}
   \right]^{-1} \, ,\\
   (\Sigma_S^{t+1})^2  &=& \left[\pi
     \frac{\overline{a^2}^t}{\overline{(y-\omega)^2}^{t+1}}
   \right]^{-1} \, ,\\
R_{il}^{t+1} &=& a_{il}^t - a_{il}^t
\frac{\overline{s}^t}{\overline{r^2}^t} + \frac{  \sum_{\nu}  (y_{\nu l} -
  \omega_{\nu l }^{t+1}) r_{\nu i}^t }{\alpha  \overline{r^2}^t}\, ,\\
S_{\nu i}^{t+1} &=&r_{\nu i}^t - r_{\nu i}^t
\frac{\overline{c}^t}{\overline{a^2}^t} +\frac {\sum_{l} (y_{\nu l} -
  \omega_{\nu l }^{t+1}) a_{ il}^t}{N \pi \overline{a^2}^t}\, ,\\
a^{t+1}_{il} &=&   f_a\left((\Sigma_R^{t+1})^2,R^{t+1}_{il}\right) ,  \label{TAP_a}\\
v^{t+1}_{il} &=& f_c\left((\Sigma_R^{t+1})^2,R^{t+1}_{il}\right) \, , \label{TAP_v}\\
r^{t+1}_{\mu i} &=&   f_r\left((\Sigma_S^{t+1})^2,S^{t+1}_{\mu i}\right) ,  \label{TAP_r}\\
s^{t+1}_{\mu i} &=& f_s\left((\Sigma_S^{t+1})^2,S^{t+1}_{\mu
    i}\right)\, . \label{TAP_s}\eea 
where only the following functions are prior-dependent:
\bea f_a(\Sigma^2,T) &=& \frac{ \rho\, 
  e^{-\frac{T^2}{2(\Sigma^2+1)}}
  \frac{\Sigma}{(\Sigma^2+1)^{\frac{3}{2}}} (\Sigma^2 + T ) }{
  (1-\rho) e^{-\frac{T^2}{2\Sigma^2}} + \rho
  \frac{\Sigma}{\sqrt{\Sigma^2+1}}
  e^{-\frac{T^2}{2(\Sigma^2+1)}} } \, , \label{f_a}\\
f_c (\Sigma^2,T) &=& \Sigma^2 \frac{{\rm d}}{{\rm d} T} f_a
(\Sigma^2,T) \, ,\label{f_c} \\
f_r(\Sigma^2,T) &=& 
\frac{ T + \Sigma^2 F'_{\mu i} \frac{\sqrt{1+\eta}}{\sqrt{N}\eta} }{(1+ \frac{1}{\eta})\Sigma^2 + 1   } \, , \\
f_s(\Sigma^2,T) &=& \frac{1}{N} \frac{ \Sigma^2 }{(1+
  \frac{1}{\eta})\Sigma^2 + 1} \, .\eea
Initial conditions are set so that the marginals correspond to the
means and variances of the prior, and $\omega_{\mu l}=y_{\mu l}$.
One iteration of the algorithm takes $O(N^3)$ steps. In
practice, we also damp the expressions
(\ref{TAP_a},\ref{TAP_v},\ref{TAP_r},\ref{TAP_s}) to ensure
convergence. If the matrix elements are not learned, this
algorithm reduces to the AMP for matrix uncertainty from
\cite{KrzakalaMezard13}, and is asymptotically equivalent to the
MU-AMP of \cite{SCHNITER}. The bilinear AMP suggested in
\cite{SCHNITER-BIG} for matrix decomposition consists of fixing a
current value of $F$ and its uncertainty, running MU-AMP, then fixing
a current values of $x$ and its variances, running MU-AMP, and
repeating. Whereas the difference in performances between the present
algorithm and the one of \cite{SCHNITER-BIG} is yet to be studied, it
is not clear if the later implements asymptotically the Bayes optimal
inference, and neither if the state evolution (next
paragraph) applies to it.

\subsection{State evolution} 
The AMP approach is amenable to asymptotic ($N\to \infty$) analysis
using a method known as ``cavity method'' in statistical physics
\cite{MezardParisi87b}, or ``state evolution'' in compressed sensing
\cite{DonohoMaleki09}. Given the parameters $\rho$, $\alpha$,$\pi$
$\eta$, $\Delta$, the MSE given by our approach follows, in the
infinite size limit: \bea E^{t+1} &=& (1-\rho)\int {\cal D} z f_c(\hat
m_x, z \sqrt{\hat m_x}) \nonumber \\ &+& \rho \int {\cal
  D} z   f_c(\hat m_x, z \sqrt{\hat m^2_x + \hat m_x}) \, ,\label{evolution1}\\
D^{t+1} &=& \frac{1}{\hat m_F + \frac{1+\eta}{\eta}} \,
,\label{evolution2} \eea where $\hat m_F^t$ and $\hat m_x^t$ follow
from eq.~(\ref{hat_eq}), with $E^{t}, D^t$ on the right hand side,
$E^{t=0}=\rho$, $D^{t=0}=1$, ${\cal D} z$ is a Gaussian integral, and
$f_c$ is defined by eq.~(\ref{f_c}).

From eqs.~(\ref{evolution1},\ref{evolution2}), one can show that the
evolution of the algorithm is equivalent to a steepest ascent of the
potential $\phi(E,D)$ obtained in eq.~(\ref{replica}). This explains
the peculiar meaning of the spinodal transition arising at $\pi^s$ and
shows that for $\pi > \pi^s$ our algorithm should approximates correctly the Bayes
optimal inference for matrix decomposition for large systems, $N\to
\infty$, as AMP does for compressed sensing. Note that in the later
case, the state evolution approach has been proveen rigorously
\cite{BayatiMontanari10} and would be interesting to see if this could
be generalize to the present, arguably more complex, case.

\subsection{Numerical tests} 
\begin{figure}[!t]
\centering
\includegraphics[width=3.2in]{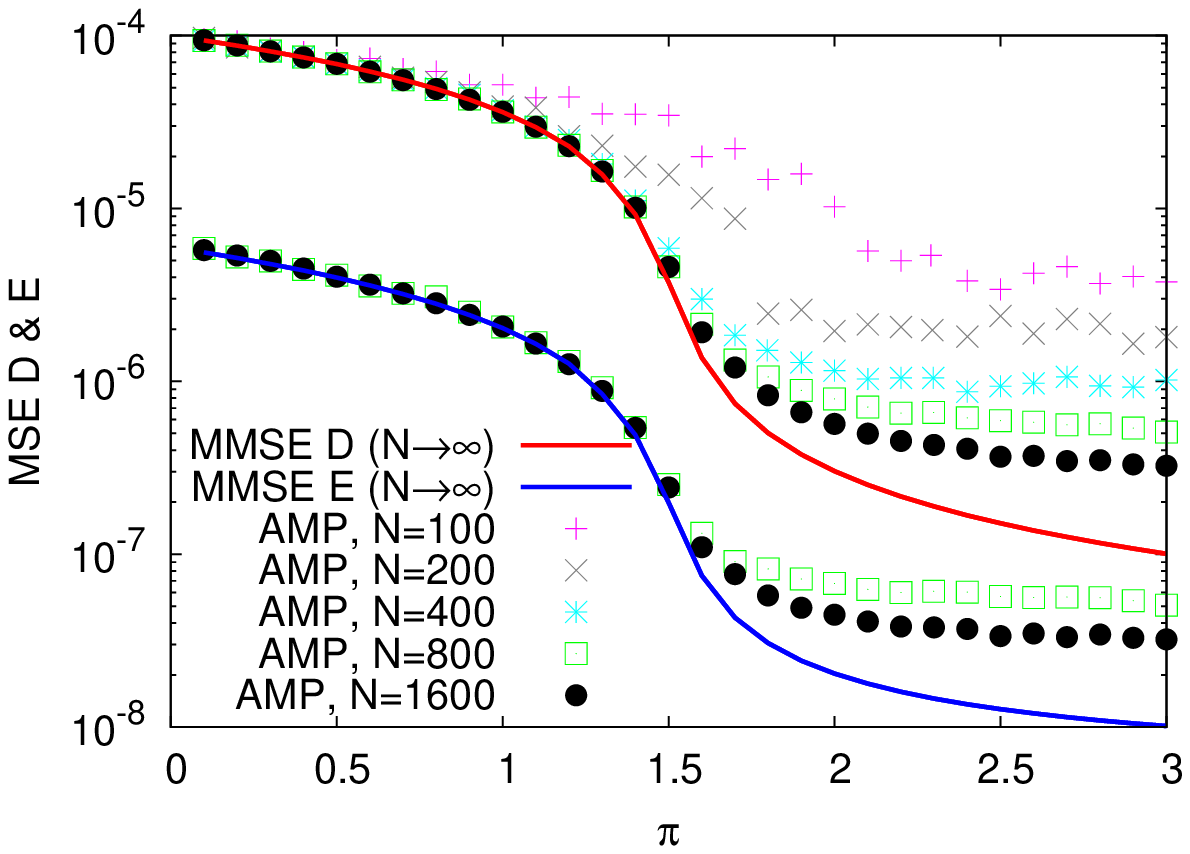}
\includegraphics[width=3.2in]{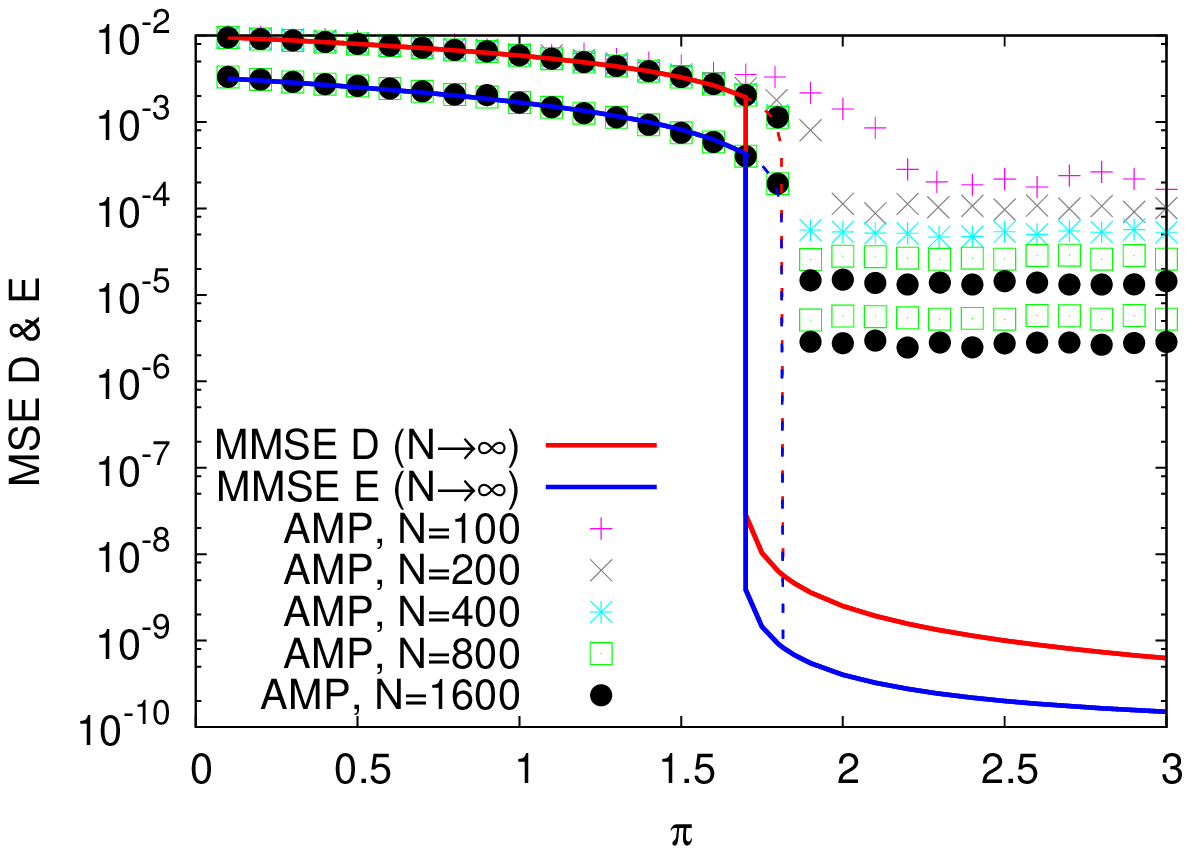}
\caption{Comparison of the performance of the AMP algorithm with the
  MMSE $D$ (for the matrix) and $E$ (for the signal, fewer points are
  shown for visibility) for different system sizes $N$. Top: A case
  with continuous decay of the MMSE for $\Delta=10^{-8}$,
  $\eta=10^{-4}$, $\alpha=0.3$, $\rho=0.1$. The initial error on the
  matrix is about $1\%$. Bottom: A case with a
  jump in the MMSE for $\Delta=10^{-8}$, $\eta=10^{-2}$, $\alpha=0.5$,
  $\rho=0.2$, the initial error on the matrix is about $10\%$. As $N$ increases, the MSE found by the algorithm approaches
  the MMSE computed theoretically. In the large $\pi$
  region, corrections to the asymptotic behavior are roughly proportional to $\eta/N$.
  \label{bpplot}}
\end{figure}

We have tested our algorithm on instances of tractable sizes. The
results are shown in Fig.~\ref{bpplot} for two noisy cases of matrix
calibration. The agreement with the
theoretical large $N$ prediction is excellent for small $\pi$. In the
large $\pi$ region, however, we observe finite-size corrections going
roughly as $\eta/N$. Despite these finite size effect, the MSE reached
by the algorithm is excellent in both case.  To appreciate the
performance of the algorithm, note that a $\ell_1$-minimization
would give very poor results even with a perfectly known matrix, as
the values of $\alpha$ and $\rho$ are above the Donoho-Tanner
transition \cite{Donoho05072005}.

The presence of $O(\eta/N)$ corrections prevents us from using our
algorithm successfully for large $\eta$. This means that so far we are
not able to solve efficiently the dictionary learning. More work will be needed to reduce
these effects.

\section{Perspectives}
It would be interesting to see if our result for the MMSE and the exactness of the state
evolution can be proven rigorously, as in compressed sensing
\cite{WuVerdu11,BayatiMontanari10}. Further, it is important to
investigate if our algorithm can be improved and the finite size
effects reduced. One can also generalize our
approach to other matrix decomposition problems and their
applications.

\section*{Acknowledgment}
After completing this study, we became aware that
\cite{SakataKabashimaNew} also analyzed the Bayes-optimal inference
for dictionary learning and reached the same conclusions as in our
Sec.~\ref{sec:replica} for $\eta\to \infty$.

This work has been supported in part by the ERC under the European
Union’s 7th Framework Programme Grant Agreement 307087-SPARCS, by the
EC Grant ‘‘STAMINA’’, No. 265496, and by the Grant DySpaN of
‘‘Triangle de la Physique.’’

\bibliographystyle{IEEEtran}
\bibliography{refs}

\end{document}